# A quick guide for student-driven community genome annotation

Authors:
Prashant S. Hosmani[1], Teresa Shippy[2], Sherry Miller[2], Joshua B. Benoit[3], Monica Munoz-Torres[4], Mirella Flores[1], Lukas A. Mueller[1], Helen Wiersma-Koch[5], Tom D'elia[5], Susan J. Brown[2] and Surya Saha[1]

[1]Boyce Thompson Institute, Ithaca, NY 14853, [2]Division of Biology, Kansas State University, Manhattan, KS 66506, [3] Department of Biological Sciences, University of Cincinnati, Cincinnati, OH 45221, [4]Lawrence Berkeley National Laboratory, Environmental Genomics and Systems Biology, Berkeley, CA 947204, [5]Indian River State College, Fort Pierce, FL 34981.

## Abstract

High quality gene models are necessary to expand the molecular and genetic tools available for a target organism, but these are available for only a handful of model organisms that have undergone extensive curation and experimental validation over the course of many years. The majority of gene models present in biological databases today have been identified in draft genome assemblies using automated annotation pipelines that are frequently based on orthologs from distantly related model organisms and usually have minor or major errors. Manual curation is time consuming and often requires substantial expertise, but is instrumental in improving gene model structure and identification. Manual annotation may seem to be a daunting and cost-prohibitive task for small research communities but involving undergraduates in community genome annotation consortiums can be mutually beneficial for both education and improved genomic resources. We outline a workflow for efficient manual annotation driven by a team of primarily undergraduate annotators. This model can be scaled to large teams and includes quality control processes through incremental evaluation. Moreover, it gives students an opportunity to increase their understanding of genome biology and to participate in scientific research in collaboration with peers and senior researchers at multiple institutions.

## Introduction

This guide describes the workflow for a community genome annotation project that connects undergraduate students with bioinformaticians, faculty and peer mentors to foster educational development and produce quality student-driven annotation. In this guide, annotation or curation is defined as the manual improvement of computationally-predicted gene structure and function associated with a genome. When annotation is integrated into undergraduate education, students get an authentic research experience and the opportunity to contribute to scientific discovery. Instructors can scale annotation projects according to the number of students and learning outcomes. The low cost of adding web-based annotation exercises to existing genome-based courses makes it an attractive option, especially for smaller institutions with limited budgets. There is substantial evidence that undergraduate research experiences that include gene annotation not only promote better understanding of genetics but also produce annotations that greatly improve existing resources [1–5]. Successful examples include large

A quick guide for student-driven community genome annotation scale, course-based programs that feature annotation as a key component [3,6]. There are many benefits for students who participate in these programs, including the development of a better understanding of genomics [3,7], retention and persistence in the sciences [6,8], and inclusion on peer-reviewed publications[9]. Bioinformatics skills required for detailed analysis of gene families and pathways align with curriculum guidelines established for biotechnology-based training [10,11]. The manual curation of gene families promotes a deeper understanding of underlying biology, a critical learning experience for undergraduate students. This provides a means to reinforce important analytical practices and standards of rigor that complement textbook learning to enhance the student's overall education. Annotation also introduces undergraduate students to the challenges of conducting research in non-model systems. Students are routinely presented with problems that are unique to the organism, gene family or pathway they are annotating. These obstacles foster the development of critical thinking, problem solving skills and resilience in a technology-based atmosphere. Students also learn to adapt to the complexities of collaborative research, a challenge that requires students to hone their teamwork and communication skills. Opportunities also exist to give students a positive research experience with sense of ownership through presentations at scientific meetings and contribution to publications.

We describe a workflow to establish curation resources, train undergraduate students, curate gene families, perform quality control and finally publish the results. The success of this community annotation model is based on a roadmap that includes building a collaborative ecosystem (see 1.1-3), recruiting new students (see 2.1) and providing them with initial training followed by continued support (see 2.3 and 3.1). The student-generated gene annotations are then reviewed before changes are committed to the official gene set. Following completion of the annotation and quality review, the student-driven annotation data is compiled for publication (see 3.3). We utilized these methods to establish the first official gene set for the insect vector of citrus greening disease [9]. Supplementary table 1 contains a list of infrastructure and resources required for manual gene annotation projects of varying scope. When successfully implemented, the student-driven annotation community model outlined here provides significant undergraduate-based educational opportunities that will yield a well-trained student population and also provide the scientific research community with quality curated data sets.

1. Build a collaborative ecosystem

1.1.    Build an ecosystem that provides supporting resources and an integrated toolkit

Manual curation is a time intensive process, but its efficiency can be improved by providing the annotators with a solid foundation of supporting resources. Most importantly, a high-quality error-corrected and near-complete genome assembly will enable the annotators to identify and correct inaccurate gene models generated by automated prediction software. The completeness of the genome assembly can be evaluated using BUSCO, a popular metric based on the expected presence of a phylogenetically appropriate set of single copy genes [12]. Validating the quality of the assembly, on the other hand, is more difficult. Alignment of paired-end RNAseq or DNAseq reads to the genome can be used to evaluate assembly errors. In general, the concordant mapping rate is an indicator of assembly quality and can be used to detect misassembled regions.

The goal of manual curation is to utilize different sources of evidence to produce the most accurate gene model possible. Automated gene prediction software, such as the MAKER annotation pipeline [13], provide a good starting point by producing consensus gene models that can be refined during manual curation. MAKER uses raw data from a variety of sources, including output from multiple *ab-initio* gene predictors, RNAseq reads and homologous proteins aligned to the genome. These evidence sources can also be used in the manual annotation process. Table 1 describes the utility of these and other resources. These resources can be prioritized based on funding and the goals of the annotation project.

| Type of data | Application |
| --- | --- |
| DNAseq | Aligned reads can help to evaluate integrity of the assembly and correct SNPs and insertion or deletion errors |
| Consensus gene predictions (e.g. MAKER [13], Prokka [14]) | Primary source of gene models for manual curation |
| Models from *ab-initio* gene prediction tools (e.g. Augustus [15], SNAP [16]) | Alternative sources of gene models that are more comprehensive but may contain false positives |
| RNAseq | Illumina short reads aligned to the genome can act as raw data for curation. They provide evidence for splicing and exon structure. RNAseq data from different tissues, organs, life stages or conditions is helpful to discern alternative transcripts. |
| Transcriptome assemblies (e.g. Trinity [17] or StringTie [18]) | These provide a condensed representation of the aligned RNAseq reads and assist in discovery of multiple isoforms. *De novo* assembled transcriptomes are a critical secondary resource to search for genes missing from the genome. |
| Homologous proteins | Well-annotated proteins from related species offer additional source evidence for validating the structure of genes. This is helpful in case of insufficient RNAseq coverage or lowly expressed genes. |

|  | Moreover, these can provide functional descriptions for the gene. |
| --- | --- |
| Full-length cDNA sequences | Pacbio or Nanopore sequencing of full-length transcripts is very useful for clearly deciphering multiple isoforms for a gene eliminating the ambiguity from partial transcripts assembled from short reads. s |
| Proteomics data | Peptides identified by mass spectrometry from different tissues of the organism can provide evidence of translation of genes predicted by *ab initio* gene predictors. |

Table 1: Use of evidence sets and other resources for manual curation.

The cost of generating each data type varies according to the genome size and other factors. Resources that can be used in multiple steps are the most cost-effective. DNAseq data used for generating the genome assembly can be reused for evaluating the assembled contigs. Paired-end RNAseq data from one or two Illumina Hiseq runs that includes multiple biological and life stages (e.g. male, female, juvenile and adult samples) can provide sufficient coverage of the gene space to be used both in an independent transcriptome assembly and for genome assembly quality checks. The RNAseq reads can also be mapped to the genome to provide evidence for expression and splicing. Pacbio Iso-Seq data is relatively expensive to produce as compared to Illumina sequence, but provides the most reliable evidence of gene structure since most reads correspond to full-length transcripts.

1.2. Identify curation targets according to project goals

Annotating all the genes in any genome assembly is a daunting task, so genes should be prioritized according to the aims of the project and available resources. Annotation can be targeted to major pathways of functional interest or gene families can be selected according to the expertise of annotators. We found it helpful for team leaders to compile an initial list of pathways or gene families to be annotated, from which students could choose genes of interest. If at all possible, members of the relevant research community should be consulted to identify the most critical targets for annotation. Not only can their interests inform the selection of genes to be annotated, but interaction with researchers who are utilizing the gene annotations helps student annotators see the significance of their work.

Once particular pathways have been selected for annotation, a metabolic pathway database can be used to classify gene families by pathway providing a useful resource for the curation effort. For example, DiaphorinaCyc [19] contains pathways for *Diaphorina citri*, the insect vector of citrus greening disease, and is used by curators for discovering genes involved in a particular pathway [20–24].

In cases where resources are available, an alternative annotation strategy is to annotate genes by walking through each scaffold [3,7,25], but this requires many annotators working simultaneously. This method also complicates background research, since investigating genes whose only connection is their position on a chromosome is much less efficient than researching gene families or metabolic pathways. Moreover, the results are not conducive to subsequent analyses, as the genes annotated likely have very little functional relatedness.

1.3. Tools for collaboration

Manual curation tends to involve teams from multiple institutions located in different parts of the country or even the world. This creates a need for robust and user-friendly platforms for collaboration and frequent meetings. Guidelines for establishing effective bioinformatics communities strongly emphasize the importance of communication and openness [26].

Various platforms are available for working in a collaborative virtual environment and should be selected based on features, previous user experience and cost. Apollo [27] is one of the most commonly used web-based manual curation tools and is open-source. There are offline gene curation tools like Artemis [28] that are also popular, but may be limited by lack of functions required for collaboration. Selecting an annotation tool with an active user community is useful for getting support from others who have faced similar issues in the past.

A majority of the communication needs can be met by using a combination of free online tools provided by Google, file sharing services like Dropbox, and a Wiki site. However, coordinating a large team of annotators located in different organizations may require project management solutions such as Basecamp, Atlassian Confluence or Asana that offer a common interface for managing multiple projects, user access, file sharing and forums. These may be available for free to educational institutions in some cases. Video conference platforms such as Google Hangouts, Skype and Zoom can be utilized for meetings.

2. Train annotators and formalize curation practices

2.1. Recruiting annotators - harnessing the crowd

Undergraduate annotators can be recruited by offering annotation as part of their coursework. An effective strategy is to use annotation as part of a capstone or senior research experience, which gives students enough time to learn annotation skills along with an incentive to complete their analysis and provide a thorough report of their gene family or pathway. A sustainable program introduces annotation over the period of a semester and then utilizes a second semester or summer for completion of gene annotation, comparative analysis and writing of final reports. This expanded time frame helps to create experienced annotators who can mentor the next cohort of students. Additional incentives for undergraduate students include the potential to present their work at scientific meetings and to contribute to peer-reviewed publications. These motivating factors help recruit responsible students who are interested in research as they begin to consider their post-undergraduate career and education options. Depending on funding, paid internships for students to participate in annotation also assists with recruitment. Overall, providing a research project focused on learning coupled with opportunities to publish builds a sense of ownership, reduces student turnover and ensures high quality annotation by undergraduate students.

2.2. Build teams according to expertise and annotation targets

Building teams of annotators has proven to be successful in our experience. If the annotation group is just starting, it is important to have a mix of undergraduate and graduate students or postdocs where the senior personnel help in defining annotation goals. At a minimum, undergraduate student participants should have completed an introductory course in genetics with an emphasis on molecular biology. Prior sequence analysis or familiarity with genome sequencing is not required and should not be a prerequisite for participation. Depending on the level of instructional support available, completion of more advanced coursework in bioinformatics may be helpful in reducing the learning curve. However, regardless of the students' educational background, the annotation process provides an effective means for students to expand their understanding of molecular and genome biology. Teaming group leaders, experienced annotators and new annotators creates an environment that fosters learning and produces quality annotation. If sufficient numbers of students are available, it might be worthwhile to implement gamification and set up competing teams that can motivate each another to curate more gene models. This has been applied in the CACAO community functional curation project ([https://gowiki.tamu.edu/wiki/index.php/Category:CACAO](https://gowiki.tamu.edu/wiki/index.php/Category:CACAO)).

2.3. Train annotators and start curation

Student-driven annotation requires initial introduction to the annotation process and continued educational instruction within a framework of peer support. In-person workshops and webinars are effective for initial training and providing students with an overview of the annotation protocols. Continued instruction is necessary to explain the detailed biology related to gene structure and interpretation of the evidence required to make more complex decisions for structural annotation. Having experienced annotators develop tutorial resources (such as PowerPoint presentations or workbook style protocols) is also very helpful for new annotators. Supplementary table 2 contains a list of free online training resources and guidelines for genome annotation. A peer mentoring network that connects new and experienced annotators is a valuable means to instruct new students and encourage teamwork. Defining regular meeting times and weekly objectives further increases the benefits of peer mentoring. Ultimately, the process of annotation is best instructed using active learning strategies. Live annotation by students in weekly meetings with group leaders, or through video conferences, is useful in getting students over hurdles they encounter during annotation. Solving these issues in a live group setting makes students more comfortable with the annotation process and provides the foundation of knowledge required to solve problems on their own.

2.4. Establish the protocols for curation

The workflow of the annotation project greatly depends on the available resources. After meeting minimum requirements of data and tools as described in the previous sections, stipulating detailed procedures and minimum standards of annotation will guide the novice curators. Annotation procedures followed by expert annotators vary based on personal preferences and also change according to the gene family under consideration. In our recent publication [9], we defined a project-specific annotation workflow that has been generalized in Figure 1 for a broader audience. Despite the potential differences in protocols followed by each annotator, we recommend that minimum evidence such as RNAseq and ortholog support are required for all manually curated genes. An evaluation process should be established (see 3.1) to ensure that these criteria are met.

Annotation workflows can be broadly grouped into three sections (i) Obtaining orthologs from closely related species, (ii) Curating gene models in an annotation editor like Apollo and (iii) Reporting the structural and functional annotation in the form of gene family or pathway reports culminating in an official gene set for the organism. Obtaining well-curated orthologs from model species will aid in structural curation. It is helpful to provide annotators with a list of closely related organisms with good quality genomic resources, from which they can collect orthologs. At this stage, a thorough literature review is recommended to gain a better understanding of the gene family or pathway and gather information relevant to the specific genes being annotated. In particular, reports of changes in gene copy number or domain organization during evolution should be noted. Student annotators should be instructed on how to keep a detailed record of their work in a lab notebook. This record can be as simple as log entries kept in a word processing document that features a cloud-based backup. Indeed, all results should be saved using a cloud-based service to prevent loss. Updates to the lab notebook should be monitored regularly, if possible, and completion should be encouraged for continuing annotation. This documentation will be essential for writing gene reports later (see 3.3).

Orthologs from related species can be used to identify candidate gene models using the BLAT [29] sequence search tool in the Apollo genome annotation editor or by BLAST [30] to organism-specific databases. Reciprocal BLAST searches should be used to verify orthology. Gene models can then be refined using available evidence tracks. Table 1 summarizes use of different evidence tracks for structural curation of gene models. The accuracy of automatically predicted and manually annotated gene models depends on the quality of the genome assembly. If the genome is highly fragmented, a *de novo* transcriptome should be used to independently validate the gene models for both structure and presence or absence. The annotators should be aware that transcriptome assembly from short read data may also produce spurious and partial transcripts.

A quick guide for student-driven community genome annotation

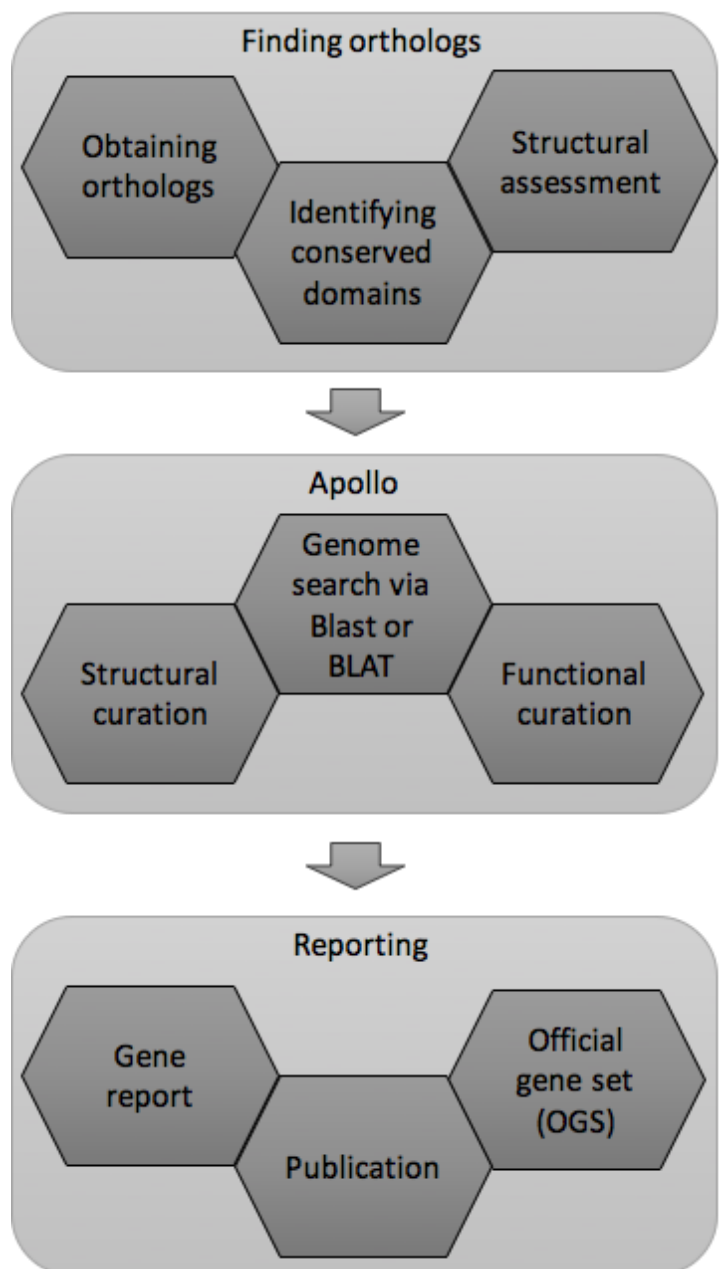

Figure 1: Annotation workflow describing various steps in manual curation of protein-coding genes.

3. Quality evaluation and publication

3.1. Iterative evaluation through peer and expert review to improve annotation

We propose a curation procedure with multiple rounds of error-checking and evaluation since the annotations are primarily performed by annotators in-training. Curation of gene models that are correctly predicted by automated gene callers and have well-curated orthologs is not prone to errors. However, lack of consensus from multiple sources of evidence and misassemblies in the genome can complicate the structural and functional annotation of a gene model. It is advantageous to have student's present updates, even if they are minor, at regular coordination meetings, so that any challenges are identified early and the students can make steady progress. Refining the annotated gene models in consultation with peers and if available, senior scientists is a useful learning experience for the students. Similar review processes can also be implemented for gene family reports, where undergraduate students evaluate each other's reports before they are presented to the senior scientists. In some cases, there may not be sufficient evidence for even expert annotators to make an informed decision about a gene model. We advise that these models be deemed putative or partial and detailed documentation kept, so that they can be resolved once new evidence or an improved genome assembly is available. The process of manual curation has been divided into specific tasks (Table 2) for each step of annotation according to Figure 1. This list can be used by instructors for grading the students through the course. We have also included an example concept inventory test (Supplementary file 1) to evaluate students after the annotation course.

| | | **Objective** | **Assessment types and descriptions** |
|---|---|---|---|
| **Finding Orthologs** | 1. | **Obtain orthologs**: Collect orthologous sequences from appropriate organisms. | Electronic lab book documentation of notes and work describing: |
| | 2. | **Identification of conserved domains**: Use online tools and databases to identify conserved domains. | 1. Names, organisms and accession numbers of orthologous sequences. Include database where orthologous sequences were collected. |
| | 3. | **Structural assessment**: Evaluate the structural organization of the gene of interest and copy numbers in closely related organisms. | 2. Names of conserved domains, size and organization within protein. Record bioinformatics tools and database used to analyze domains. |
| | | | 3. Structural organization of the gene and copy number in closely related organisms. |
| | | | 1-3. Prepare a short report (PowerPoint or written) of the gene family/pathway, share with lab group or peers in class. Reports should include: literature review and determination of gene family/pathway function, copy number of genes, conservation in related organisms, estimation of number of each gene expected to be in the family/pathway. |
| **Apollo** | 1. | **Genome Search via BLAST or BLAT:** Perform and interpret BLAT results from Apollo or BLAST results of databases. | Electronic lab book documentation of notes and work describing: |
| | 2. | **Structural Curation**: Complete manual structural annotation using evidence tracks in Apollo. | 1. Details of BLAT or BLAST results, including: Similarity or identity scores, E values, query coverage and genome coordinates of matching sequences. |
| | 3. | **Functional Curation:** Propose the functional classification of the annotated gene and show support from comparative analysis with homologous sequences, identification of conserved domains, phylogenetic analyses or other lines of evidence. | 2. Record status of predicted models and evidence tracks for gene to be annotated. Record changes made to predicted model. Evaluate structural annotation by comparison of final sequence to orthologs and data collected on conserved domains to determine the completeness of the annotation. |
| | | | 3. Document comparative analysis to homologous proteins that supports the functional characterization. Record organisms, accession numbers and sequence similarity. Provide results of analyses using BLAST, multiple sequence alignments or phylogenetic analysis. |
| | | | 1-3. Iterative annotation with review**:** Examine accuracy of annotations through peer review and presentation of short reports (PowerPoint or written) to faculty and scientist mentors. |

| | | |
|---|---|---|
| **Reporting** | 1. **Gene/pathway report**: Compose a written report to justify accuracy of final annotation and to detail results of the completed annotation in an evolutionary and genomic context.<br>2. **Official gene set:** Report data to lead scientists for official gene set.<br>3. **Publication:** Assemble reports and summaries of annotation and comparative analysis for peer-reviewed publication. | 1. Written report, poster presentations or oral presentations (class or professional meetings) that include<br>   a. Overview of gene family/pathway<br>   b. Description of the annotated genes, processes used, support and evidence collected<br>   c. Gene copy tables for each gene in family/pathway<br>   d. Pairwise comparisons of genes in other organisms<br>   e. Phylogenetic trees of genes with sequence/copy number different from those in orthologs<br>   f. Analysis of biological significance of genes in family/pathway based on evidence from related organisms<br>2. Contribute information required for establishing the official gene set.<br>3. Contribute reports and information required for preparing peer-reviewed publications. |

Table 2: Assessment plan for students with description of student objectives and related assessments to measure student annotation progress and quality. Objectives are outlined to ensure students follow the workflow in Figure 1. Students should be able to perform the activities at each step before starting the next phase of the workflow. Objective numbers correspond to the appropriate assessment type and descriptions.

3.2. Finalize annotation and evaluate quality of entire official gene set

Manually curated genes should be merged with the models from automated gene predictors after each round of annotation to create the official gene set for public release to the research community. Curated models selected for public release should be carefully screened for errors by expert curators. Tools such as the GFF3toolkit (https://github.com/NAL-i5K/GFF3toolkit/) are useful to identify errors in the curated gene models and automate the merging process. Updates to gene annotations across annotation releases can be tracked by using unique gene identifiers and version numbers. Version numbers should be incremented only if the sequence has been modified in the new annotation. Submission to public databases like the NCBI and ENA is recommended, but the process can be time consuming. There are other options like Figshare (https://figshare.com/), Dryad (http://datadryad.org/) and Ag Data Commons (https://data.nal.usda.gov/), as well as clade-specific databases like i5k (https://i5k.nal.usda.gov/ [31]) and SOL genomics (https://solgenomics.net/ [32]).

Metrics for measuring improvements over the entire gene set are limited by the availability of gold standard annotations for comparison and the inherent complexity of annotation. Annotation Edit Distance (AED) gives a measure of the transcript evidence supporting a gene model. MAKER [13] calculates the annotation edit distance for all the gene models in an annotation set. We have shown that this metric can act as a good measure for quality of annotations [9]. AED has also been adopted for model organisms like *Arabidopsis*, where it replaced the five-star based ranking method used by TAIR [33]. A genome-independent *de novo* transcriptome can also be used to validate the structure of gene models in the official gene set.

3.3. Publication

The goals for publication depend on the scope of the annotation project. Annotation followed by phylogenetic analysis and functional characterization of biologically important gene families can be presented as a course project report or a poster at scientific conferences [20–24] [https://fas.fit.edu]. Results from a larger community curation project can reflect a significant research contribution that can warrant a journal publication [25,31]. In either case, we recommend that the undergraduate annotators summarize all their findings in gene reports that can then be iteratively revised in consultation with peers and senior scientists. The gene report can be structured like a mini manuscript with an introduction followed by literature review, methods, results and discussion (See supplementary data in Saha et al., 2017 [9]). It is critical to provide uniform report guidelines, so reports can easily be merged together for publication (e.g. supplementary materials). The discussion should focus on structural features and domain organization of the gene family, in addition to copy number analysis. A phylogenetic comparison with related species can be used to provide evolutionary context to the structure and function. This exercise is helpful in training undergraduate students to present their own work and introduces them to scientific writing.

**Conclusion**

The guidelines presented here provide a framework to build a successful student-driven annotation community that can contribute to ongoing research projects. A community that consists of experts, instructors and peer mentors provides the ideal framework to train and supervise undergraduate students so they can make a meaningful contribution. Benefits to undergraduate student participants include an increase in learning, critical thinking and problem-solving abilities related to molecular biology and genomics. Community curation provides knowledge and skills that help students progress in their undergraduate courses. Moreover, a research experience also encourages exploration and pursuit of graduate education. The inherent need for communication and teamwork in a diverse and sometimes virtual community also develops skills that are transferable to a wide range of careers. Students are excited to participate in research projects that have tangible scientific outcomes. This builds a sense of ownership and responsibility, resulting in students who are eager to annotate and also mentor beginners to sustain the community. It is advisable for organizers to plan early for the turnover of annotators so there is sufficient overlap of the incoming cohort with experienced annotators.

We effectively deployed this strategy during a three-year period to train over 40 student annotators from four different institutions including three universities, a state college, and a research institute. The entire community manually curated approximately 530 gene models. Other accomplishments include creation of the first official gene set for *D. citri*, an important insect vector, and a peer-reviewed publication featuring student annotators as contributing authors [9]. These results demonstrate that the student-driven community model is fully capable of producing high quality gene models while providing a supportive and valuable educational experience for undergraduate students.

A quick guide for student-driven community genome annotation

**Acknowledgements**

We would like to thank Monica F. Poelchau from USDA-ARS, NAL for useful suggestions on the manuscript. All open-access fees, student annotators and post-docs were funded through USDA-NIFA grant 2015-70016-23028.

**References**

1. Buckner B, Beck J, Browning K, Fritz A, Grantham L, Hoxha E, et al. Involving Undergraduates in the Annotation and Analysis of Global Gene Expression Studies: Creation of a Maize Shoot Apical Meristem Expression Database. Genetics. 2007;176.
2. Mitchell CS, Cates A, Kim RB, Hollinger SK. Undergraduate Biocuration: Developing Tomorrow's Researchers While Mining Today's Data. J Undergrad Neurosci Educ. Faculty for Undergraduate Neuroscience; 2015;14: A56-65. Available: http://www.ncbi.nlm.nih.gov/pubmed/26557796
3. Shaffer CD, Alvarez C, Bailey C, Barnard D, Bhalla S, Chandrasekaran C, et al. The genomics education partnership: successful integration of research into laboratory classes at a diverse group of undergraduate institutions. CBE Life Sci Educ. American Society for Cell Biology; 2010;9: 55–69. doi:10.1187/09-11-0087
4. Beagley CT. Genome annotation in a community college cell biology lab. Biochem Mol Biol Educ. Wiley-Blackwell; 2013;41: 44–49. doi:10.1002/bmb.20669
5. Pope WH, Bowman CA, Russell DA, Jacobs-Sera D, Asai DJ, Cresawn SG, et al. Whole genome comparison of a large collection of mycobacteriophages reveals a continuum of phage genetic diversity. Elife. 2015;4. doi:10.7554/eLife.06416
6. Jordan, T. C., Burnett, S. H., Carson, S., Caruso, S. M., Clase, K., DeJong, R. J., Dennehy, J. J., Denver, D. R., Dunbar, D., Elgin, S. C. R., Findley A. M., Gissendanner, C. R., Golebiewska, U. P., Guild, N., Hartzog, G. A., Grillo, W. H., Hollowell, G. GF. A broadly implementable research course for first-year undergraduate students. MBio. 2014;In Press: 1–8. doi:10.1128/mBio.01051-13.Editor
7. Shaffer CD, Alvarez CJ, Bednarski AE, Dunbar D, Goodman AL, Reinke C, et al. A course-based research experience: how benefits change with increased investment in instructional time. CBE Life Sci Educ. American Society for Cell Biology; 2014;13: 111–30. doi:10.1187/cbe-13-08-0152
8. Hanauer DI, Graham MJ, Betancur L, Bobrownicki A, Cresawn SG, Garlena RA, et al. An inclusive Research Education Community (iREC): Impact of the SEA-PHAGES program on research outcomes and student learning. Proc Natl Acad Sci. 2017; 201718188. doi:10.1073/pnas.1718188115
9. Saha S, Hosmani PS, Villalobos-Ayala K, Miller S, Shippy T, Flores M, et al. Improved annotation of the insect vector of citrus greening disease: biocuration by a diverse genomics community. Database. 2017;2017: bax032. doi:10.1093/database/bax032
10. White HB, Benore MA, Sumter TF, Caldwell BD, Bell E. What skills should students of undergraduate biochemistry and molecular biology programs have upon graduation? Biochem Mol Biol Educ. 2013;41: 297–301. doi:10.1002/bmb.20729
11. Welch L, Lewitter F, Schwartz R, Brooksbank C, Radivojac P, Gaeta B, et al. Bioinformatics Curriculum Guidelines: Toward a Definition of Core Competencies. PLoS Comput Biol. Public Library of Science; 2014;10: e1003496. doi:10.1371/journal.pcbi.1003496
12. Simao FA, Waterhouse RM, Ioannidis P, Kriventseva E V., Zdobnov EM. BUSCO:

assessing genome assembly and annotation completeness with single-copy orthologs. Bioinformatics. 2015;31: 3210–3212. doi:10.1093/bioinformatics/btv351
13. Cantarel BL, Korf I, Robb SMC, Parra G, Ross E, Moore B, et al. MAKER: an easy-to-use annotation pipeline designed for emerging model organism genomes. Genome Res. 2008;18: 188–96. doi:10.1101/gr.6743907
14. Seemann T. Prokka: rapid prokaryotic genome annotation. Bioinformatics. 2014; btu153-. doi:10.1093/bioinformatics/btu153
15. Stanke M, Keller O, Gunduz I, Hayes A, Waack S, Morgenstern B. AUGUSTUS: A b initio prediction of alternative transcripts. Nucleic Acids Res. 2006;34. doi:10.1093/nar/gkl200
16. Korf I. Gene finding in novel genomes. BMC Bioinformatics. 2004;5. doi:10.1186/1471-2105-5-59
17. Grabherr MG., Brian J. Haas, Moran Yassour Joshua Z. Levin, Dawn A. Thompson, Ido Amit, Xian Adiconis, Lin Fan, Raktima Raychowdhury, Qiandong Zeng, Zehua Chen, Evan Mauceli, Nir Hacohen, Andreas Gnirke, Nicholas Rhind, Federica di Palma, Bruce W. N, Friedman and AR. Trinity: reconstructing a full-length transcriptome without a genome from RNA-Seq data. Nat Biotechnol. 2013;29: 644–652. doi:10.1038/nbt.1883.Trinity
18. Pertea M, Pertea GM, Antonescu CM, Chang T-C, Mendell JT, Salzberg SL. StringTie enables improved reconstruction of a transcriptome from RNA-seq reads. Nat Biotechnol. Nature Publishing Group, a division of Macmillan Publishers Limited. All Rights Reserved.; 2015; 290–295. doi:10.1038/nbt.3122
19. Flores M, Saha S, Hosmani PS, Brown S, Mueller LA. Community portal for the Citrusgreening disease resources. 2017; doi:10.6084/m9.figshare.5307391.v1
20. Vosburg C, Wiersma-Koch H, D'Elia T. Annotation of Segment Polarity genes in Diaphorina citri. Florida Sci. 2018;81:S1.
21. Grace R, Wiersma-Koch H, D'Elia T. Genomic identification, annotation and comparative analysis of Vaculolar-type ATP synthase subunits in Diaphorina citri. Florida Sci. 2018;81:S1.
22. Villalobos-Ayala K, Cordola C, Bell T, Wiersma-Koch H, Hunter W, D'Elia T. Genomic analysis of Diaphorina citri, Asian citrus psyllid reveals numerous classes of heat shock proteins. Florida Sci. 2016;79:5-6.
23. Bell T, Cordola C, Villalobos-Ayala K, Wiersma-Koch H, Hunter W, D'Elia T. Analysis and characterization of the cathepsin gene family in the Asian citrus psyllid, Diaphorina citri (Hemiptera: Liviidae). Florida Sci. 2016;
24. Cordola C, Villalobos-Ayala K, Bell T, Wiersma-Koch H, Hunter W, D'Elia T. Identification and analyses of Rab genes in the genome of Asian citrus psyllid (Hemiptera: Liviidae). Florida Sci. 2016;79:6-7.
25. Leung W. Drosophila Muller F elements maintain a distinct set of genomic properties over 40 million years of evolution. G3. 2015;63. doi:10.1534/g3.114.015966
26. Budd A, Corpas M, Brazas MD, Fuller JC, Goecks J, Mulder NJ, et al. A Quick Guide for Building a Successful Bioinformatics Community. PLoS Comput Biol. 2015;11: e1003972. doi:10.1371/journal.pcbi.1003972
27. Lee E, Helt G, Reese J, Munoz-Torres MC, Childers C, Buels RM, et al. Web Apollo: a web-based genomic annotation editing platform. Genome Biol. 2013;14: R93. doi:10.1186/gb-2013-14-8-r93
28. Carver T, Harris SR, Berriman M, Parkhill J, McQuillan JA. Artemis: An integrated platform for visualization and analysis of high-throughput sequence-based experimental data. Bioinformatics. 2012;28: 464–469. doi:10.1093/bioinformatics/btr703
29. Kent WJ. BLAT — The BLAST -Like Alignment Tool. Genome Res. 2002;12: 656–664. doi:10.1101/gr.229202.
30. Altschul SF, Gish W, Miller W, Myers EW, Lipman DJ. Basic local alignment search tool.

1990. pp. 403–410.
31. Poelchau M, Childers C, Moore G, Tsavatapalli V, Evans J, Lee C-Y, et al. The i5k Workspace@NAL--enabling genomic data access, visualization and curation of arthropod genomes. Nucleic Acids Res. 2014;43: D714–D719. doi:10.1093/nar/gku983
32. Fernandez-Pozo N, Menda N, Edwards JD, Saha S, Tecle IY, Strickler SR, et al. The Sol Genomics Network (SGN)-from genotype to phenotype to breeding. Nucleic Acids Res. 2014; gku1195-. doi:10.1093/nar/gku1195
33. Cheng C-Y, Krishnakumar V, Chan A, Schobel S, Town CD. Araport11: a complete reannotation of the Arabidopsis thaliana reference genome [Internet]. bioRxiv. Cold Spring Harbor Labs Journals; 2016 Apr. doi:10.1101/047308